\documentclass[a4paper,11pt]{article}

\usepackage{jheppub} 

\usepackage[T1]{fontenc} 
\usepackage{color,graphicx,epsfig}

\title{\boldmath Unsupervised quark/gluon jet tagging with Poissonian Mixture Models}

\author[a]{E.~Alvarez,}
\author[b,c]{M.~Spannowsky}
\author[a]{and M.~Szewc}

\affiliation[a]{International Center for Advanced Studies (ICAS) and CONICET, UNSAM,
Campus Miguelete, 25 de Mayo y Francia, CP1650, San Martin, Buenos Aires, Argentina}
\affiliation[b]{Institute for Particle Physics Phenomenology, 
Durham University, DH1 3LE, United Kingdom}
\affiliation[c]{Department of Physics,
Durham University, DH1 3LE, United Kingdom}

\emailAdd{sequi@unsam.edu.ar}
\emailAdd{michael.spannowsky@durham.ac.uk}
\emailAdd{mszewc@unsam.edu.ar}

\preprint{IPPP/21/58 \& ICAS 70/21}

\abstract{The classification of jets induced by quarks or gluons is important for New Physics searches at high-energy colliders. However, available taggers usually rely on modelling the data through Monte Carlo simulations, which could veil intractable theoretical and systematical uncertainties.  To significantly reduce biases, we propose an unsupervised learning algorithm that, given a sample of jets, can learn the SoftDrop Poissonian rates for quark- and gluon-initiated jets and their fractions.  We extract the Maximum Likelihood Estimates for the mixture parameters and the posterior probability over them.  We then construct a quark-gluon tagger and estimate its accuracy in actual data to be in the $0.65-0.7$ range, below supervised algorithms but nevertheless competitive. We also show how relevant unsupervised metrics perform well, allowing for an unsupervised hyperparameter selection. Further, we find that this result is not affected by an angular smearing introduced to simulate detector effects for central jets.  The presented unsupervised learning algorithm is simple; its result is interpretable and depends on very few assumptions.}

\begin{document} 
\maketitle
\flushbottom

\section{Introduction}\label{sec:intro}

As ongoing searches at the LHC have not succeeded in providing guidance on the nature of extensions of the Standard Model, unbiased event reconstruction and classification methods for new resonances and interactions have become increasingly important to ensure that the gathered data is exploited to its fullest extent~\cite{Kasieczka:2021xcg,Aarrestad:2021oeb}. Unsupervised data-driven classification frameworks, often used as anomaly-detection methods, are comprehensive in scope, as their success does not hinge on how theoretically well-modelled signal or background processes are allowing for more signal-agnostic analyses~\cite{Choi:2020bnf,Caron:2021wmq,Dohi:2020eda,dAgnolo:2021aun,Nachman:2020lpy,Andreassen:2020nkr,Hajer2020,roy2020robust,Andreassen:2018apy}.


Here, we apply the unsupervised-learning paradigm to the discrimination of jets induced by quarks or gluons. The so-called quark-gluon tagging of jets can be a very powerful method to separate signal from background processes. Important examples include the search for dark matter at colliders, where the dark matter candidates are required to recoil against a single hard jet \cite{CMS:2014jvv}, the measurement of Higgs boson couplings in the weak-boson fusion process \cite{Dokshitzer:1991he,Rainwater:1998kj} or the discovery of SUSY cascade decays involving squarks or gluinos \cite{Bhattacherjee:2016bpy}. Thus, a robust and reliable method to discriminate between quark and gluon jets furthers the scientific success of the LHC programme in precision measurements and searches for new physics. Consequently, several approaches have been proposed to exploit the differences in the radiation profiles of quarks and gluons \cite{Gallicchio:2011xq,Larkoski:2013eya,Larkoski:2014pca,Bhattacherjee:2015psa,FerreiradeLima:2016gcz,Kasieczka:2018lwf} and have been studied in data by ATLAS \cite{ATLAS:2014vax} and CMS \cite{CMS:2013kfa}. 

The discrimination of quarks and gluons as incident particles for a jet poses a challenging task. Some of the best performing observables to classify quark/gluon jets are infrared and/or collinear (IRC) unsafe, e.g. the number of charged tracks of a jet. Thus, evaluating the classification performance of IRC unsafe observables from the first principles is an inherently difficult task. Instead, SoftDrop~\cite{Larkoski:2014wba} has been shown to achieve a high classification performance while maintaining IRC safety. Further, at leading-logarithmic accuracy, the SoftDrop multiplicity $n_\mathrm{SD}$ exhibits a Poisson-like scaling \cite{Frye:2017yrw}, allowing us to construct an entire data-driven unsupervised classifier based on a mixture model. Although IRC safety is in principle not necessary to construct an unsupervised tagger, the fact that we are able to know the leading-logarithmic behavior of the observable is what allows us to build a simple and interpretable probabilistic model. To build a tagger that discriminates between quark and gluon jets, we extract the Maximum Likelihood Estimate (MLE) and the posterior distributions for the rate of the Poissonians and the mixing proportions of the respective classes. We find such a tagger to have a high accuracy ($\approx 0.7$) while remaining insensitive to detector effects. In a second step, we augment this method by using Bayesian inference to obtain the full set of posterior distributions and correlations between the model parameters, which allows to calculate the probability of a jet being a quark or gluon jet. The latter method results in a robust tagger with even higher accuracy. Thus, this approach opens a novel avenue to analyse jet-rich final states at the LHC, thereby increasing the sensitivity in searches for new physics.  

The structure of the paper is as follows: In section \ref{sec:method} we present the datasets considered and describe the mixture model method for the discrimination of quark and gluon jets. In section \ref{sec:results} we discuss the performance and uncertainties of the MLE algorithm, detailing the viability of this algorithm in the presence of detector effects. We use Bayesian inference to obtain the full posterior probability density function in section \ref{sec:bayes}. In section \ref{sec:outlook} we offer a summary and conclusions.

\section{Mixture models for quark- and gluon-jets data}\label{sec:method}

To showcase and benchmark our model performance in a way comparable to other algorithms, we have considered two datasets available in the literature, both considered initially in Ref.~\cite{Komiske:2018cqr}. These two datasets contain quark and gluon jets after hadronization, and correspond to a set~\cite{Zenodo:EnergyFlow:Pythia8QGs} generated with Pythia~\cite{Sjostrand:2014zea} and a set~\cite{Zenodo:EnergyFlow:Herwig7QGs} generated with Herwig~\cite{Bahr:2008pv,Bellm:2015jjp}.  The reason for using datasets from different generators is to verify that the algorithm is independent of the generator and from any specific tuning.  As detailed in the documentation, the quark- and gluon-initiated jets are generated from $qg\rightarrow Z(\rightarrow\nu\overline{\nu})+u/d/s$ and $q\overline{q}\rightarrow Z(\rightarrow\nu\overline{\nu})+g$ processes in $pp$ collisions at $\sqrt{s} $ = 14 TeV.  After hadronization, the jets are clustered using the anti-$k_{T}$ algorithm with $R=0.4$. For the sake of validation and comparing supervised and unsupervised metrics, we use the parton level information to define whether a jet is a true quark or a true gluon. This definition is known to be problematic and we emphasize that our model does not depend on these unphysical labels and could instead provide an operational definition of quark and gluon jets~\cite{Komiske:2018vkc}. In addition, there is a very strict selection cut and all the provided jets have transverse momentum $p_{T}\in[500.0,550.0]$ GeV and rapidity $|y|\, < 1.7$. We detail the impact of these cuts on the tagging observable in the following paragraphs. Finally, the dataset is balanced with an equal number of quark and gluon jets.

As a tagging observable, we have considered the Iterative SoftDrop Multiplicity $n_{\mathrm{SD}}$ defined in Ref.~\cite{Frye:2017yrw}. Once defined the jet radius $R$ used to cluster the constituents with the Cambridge-Aachen algorithm~\cite{Dokshitzer:1997in}, $n_{\mathrm{SD}}$ has three hyperparameters $z_{\text{cut}}$, $\beta$ and $\theta_{\text{cut}}$. The dependence on these hyperparameters and the classification performance on supervised tasks has been explored in Ref.~\cite{Frye:2017yrw}. In this work and in agreement with Ref.~\cite{Frye:2017yrw} we will consider IRC safe parameter choices: $z_{\text{cut}} > 0 $, $\beta < 0 $ and $\theta_{\text{cut}} = 0 $.

The choice of a well-known tagging observable allows us to perform unsupervised quark-gluon discrimination by considering interpretable mixture models~\cite{Metodiev:2017vrx, Komiske:2018oaa,Metodiev:2018ftz,Dillon:2019cqt, Alvarez:2019knh, Dillon:2020quc,Alvarez:2021hxu,Graziani:2021vai,_t_p_nek_2015,Dillon:2021aeo}, where we think of the measurement of $N$ jets and their $n_\mathrm{SD}$, values as originating from underlying themes, which we would ideally match with quark and gluon jets. In a probabilistic modelling framework, we want to obtain the underlying quark and gluon distributions from the observed data $X=\{n^{(i)}_{\mathrm{SD}},i=1,...,N\}$, which has a likelihood function
\begin{equation}
    p(X)=\prod_{i=1}^{N}p(n^{(i)}_{\mathrm{SD}})=\prod_{i=1}^{N}\sum_{k=\{q,g\}}\pi_{k}\,p(n^{(i)}_{\mathrm{SD}}\,|\,k),
\end{equation}
where $k$ is the jet class or theme. The mixing fraction $\pi_{k}$ denotes the probability of sampling a jet from theme $k$ and $p(n^{(i)}_{\mathrm{SD}}\,|\,k)$ is the $n_\mathrm{SD}$ probability mass functions conditioned on which theme the jet belongs to. In principle, the number of themes is a hyperparameter of the model and could be optimized with some criteria (see e.g. Ref.~\cite{celeux:hal-01961077} for a review on different methods to select the number of themes).  In this work, we consider only two themes that we identify with quark and gluon jets. This choice is based on physical grounds. As we will detail in the following paragraphs, for a sufficiently small $p_{T}$ range we only expect two themes. The final ingredient to build the probabilistic model is the specification of $p(n^{(i)}_{\mathrm{SD}}\,|\,k)$. To model these probability mass functions, we make use of the fact that at leading logarithmic (LL) order $n_\mathrm{SD}$ is Poisson distributed~\cite{Frye:2017yrw}:
\begin{equation}
    p(X)=\prod_{i=1}^{N}\sum_{k=\{q,g\}}\pi_{k}\,\text{Poisson}(n^{(i)}_{\mathrm{SD}};\lambda_{k}), \label{eq:mixture_model}
\end{equation}
where $\lambda_{k}$ is the Poisson rate for each theme, which fixes the mean and variance of the $n_\mathrm{SD}$ distribution. The departure of the Poisson hypothesis by NLL corrections and Non Perturbative effects can then be seen by examining the variance to mean ratio for each class of jets. We see that deviations from the Poissonian behaviour are parameter-dependent, in agreement with previous results detailed in Ref.~\cite{Frye:2017yrw}. Furthermore, we see that the deviations are more substantial for quark jets than for gluon jets. This is enhanced by the fact that quark jets usually have smaller $n_{\mathrm{SD}}$ values than gluon jets. 

The behavior of $n_\mathrm{SD}$ is also dependent on the kinematics of the jet. For the samples considered, the limited $p_{T}$ and $|y|$ ranges ensure that all quark- and gluon-initiated jets follow the same respective $n_{\mathrm{SD}}$ distributions. In a more realistic implementation of this model where the $p_{T}$ of the jets populates a much wider range, the model implementation should be modified to account for the variation of the $n_{\mathrm{SD}}$ distribution with $p_{T}$. A straightforward strategy is to bin the $p_{T}$ distribution and infer the mixture model parameters in each bin, effectively conditioning the Poisson rates and the mixture fractions on the $p_{T}$ of the jets populating such region. The $p_{T}$ dependence also implies that the discriminating power of the mixture model will depend on the $p_{T}$ of the jet as is usually the case for most quark/gluon classification methods.

\begin{figure}[ht!]
    \centering
    \includegraphics[width=\textwidth]{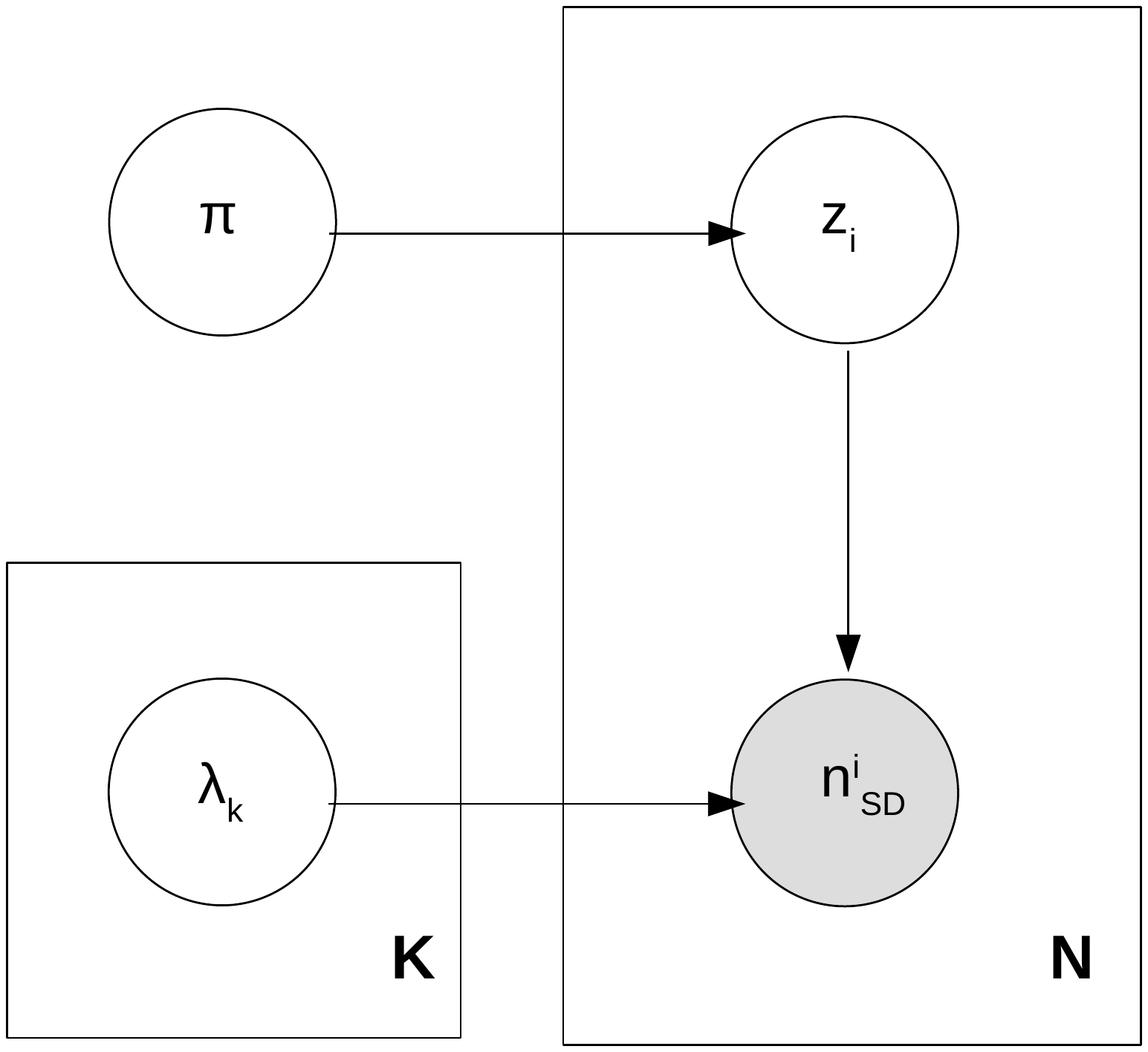}
    \caption{Generative process for $n_\mathrm{SD}$ due to a mixture of quark and gluon jets.}
    \label{fig:generative_process}
\end{figure}

The likelihood in Eq.~\ref{eq:mixture_model} describes how for a given value of the mixing fractions $\pi_{q,g}$ and the Poisson rates $\lambda_{q,g}$, each jet is sampled or generated. This is called a generative process and it is often useful to represent it as a plaque diagram~\cite{bishop}. The corresponding plaque diagram to Eq.~\ref{eq:mixture_model} can be seen in Fig.~\ref{fig:generative_process}.

In the figure, we have introduced a hidden or latent class variable, the theme assignment $z$, which dictates whether the generated jet is a quark or a gluon jet. This class assignment is necessary to think of the likelihood as a generative process and it is useful when performing inference and when building a probabilistic jet classifier. Having defined the probabilistic model and the relevant parameters $\pi$ and $\lambda$, finding the underlying themes becomes synonymous with finding the posterior probabilities for $\pi$ and $\lambda$. These posterior probabilities can then be used to build a quark/gluon classifier. Instead of tackling the Bayesian Inference problem head-on, one can first obtain point estimates for $\pi$ and $\lambda$. Because we consider a statistically significant dataset and a straightforward model, at this stage we consider the Maximum Likelihood Estimates (MLE) of $\pi$ and $\lambda$ instead of Maximum A Posterior (MAP) estimates. These estimates can be obtained easily through Expectation-Maximization (EM) or with Stochastic Variational Inference through dedicated software such as the Pyro package~\cite{bingham2018pyro,phan2019composable}. We need to be careful when estimating the point parameters as they can suffer from mode degeneracy and mode collapse. The former occurs due to the permutation symmetry of the classes and can be fixed by requiring that $\lambda_{q}<\lambda_{g}$ as dictated by basic principles. The latter occurs when one theme is emptied of samples. Because we only consider two classes, collapse is avoided for for good hyperparameter choices because of the multimodality of the data distribution.

With the MLE point estimation of $\pi^{\text{MLE}}$ and $\lambda^{\text{MLE}}$, we can construct a probabilistic jet classifier by computing the assignment probabilities or responsibilities,
\begin{equation}
    p(z=\text{quark}\,|\,n_\mathrm{SD},\pi^{\text{MLE}},\lambda^{\text{MLE}})=\frac{\pi^{\text{MLE}}_{q}\,\text{Poisson}(n_\mathrm{SD},\lambda^{\text{MLE}}_{q})}{\sum_{k=\{q,g\}}\pi^{\text{MLE}}_{k}\,\text{Poisson}(n_\mathrm{SD},\lambda^{\text{MLE}}_{k})},
    \label{eq:classifier}
\end{equation}
with $p(z=\text{gluon})=1-p(z=\text{quark})$.  The classifier is obtained by selecting a threshold $0\leq c\leq1.0$ and labeling any jet with $p(z=\text{quark}\,|\,n_\mathrm{SD},\pi^{\text{MLE}},\lambda^{\text{MLE}})\geq c$ as a quark jet. This classifier has a clear probabilistic justification and it is interpretable, which is a considerable asset for an unsupervised task. 

For validation, we compute the usual supervised metrics: the accuracy obtained by assigning classes using the probabilistic working point $c=0.5$ chosen because we have a binary classification problem and a probabilistic algorithm, the mistag rate at $50\%$ signal efficiency $\epsilon^{-1}_{g}(\epsilon_{q}=50\%)$ and the Area-Under-Curve (AUC). The accuracy is defined as the number of fraction of well classified samples, $\epsilon_{q,g}$ are the fraction of well classified quark/gluon jets and the AUC is the integral of the Receiver Operating Characteristic (ROC) curve $\epsilon_{q}(\epsilon_{g})$ with a higher AUC usually signaling a higher overall performance. However, because we are interested in an unsupervised classifier trained directly on data, we also define unsupervised metrics. These metrics need to be correlated with the unseen accuracy so as to substitute it as a measure of performance in a fully data-driven implementation of the model. In an unsupervised metric we measure how consistent is the learned model with the measured data. We investigate two metrics that encode such consistency:
\begin{eqnarray}
d_{H}(p,q)&=&\frac{1}{\sqrt{2}}\sqrt{\sum_{n_\mathrm{SD}=0}^{\infty}(\sqrt{p(n_\mathrm{SD})}-\sqrt{q(n_\mathrm{SD})})^{2}}\nonumber\\
\text{KL}(p||q)&=&-\sum_{n_\mathrm{SD}=0}^{\infty}p(n_\mathrm{SD})\text{Ln}\left(\frac{q(n_\mathrm{SD})}{p(n_\mathrm{SD})}\right),
\end{eqnarray}
where $d_{H}$ is the Hellinger distance~\cite{Deza2009} and KL is the Kullback-Leibler divergence between the learned data density and the measured data density. The latter can be interpreted as the amount of information needed to approximate samples that follow the distribution $p$ with samples generated by a model $q$. In this paper, $p$ will be the measured data density obtained by the $n_\mathrm{SD}$ frequencies and $q$ will be the posterior predictive distribution $q(n_\mathrm{SD})=\sum_{k=\{q,g\}}\pi^{\text{MLE}}_{k}\,\text{Poisson}(n_\mathrm{SD},\lambda^{\text{MLE}}_{k})$. Other metrics such as the Energy Mover's Distance~\cite{Komiske:2019fks} could also be applied. We emphasize that this takes advantage of the fact that we are learning more than a classifier, as we are modelling the data density itself and the underlying processes that generate it. If we can match the learned models to quark and gluon jets, it means we can understand the data beyond merely a good discriminator.

In section~\ref{sec:results}, we apply this model to the two quark and gluon datasets~\cite{Zenodo:EnergyFlow:Pythia8QGs},~\cite{Zenodo:EnergyFlow:Herwig7QGs} and obtain the different MLE point parameters and derived metrics for other choices of SoftDrop hyperparameters. We then go beyond the point estimate calculation by introducing priors for $\pi$ and $\lambda$ and obtain the corresponding posterior through numerical Bayesian inference in section~\ref{sec:bayes}. These priors can encode our theoretical domain knowledge, such as the LL estimates of $\lambda_{q}$ and $\lambda_{g}$ and also regularize our model and thus avoid mode degeneracy and mode collapse.

\section{Results}\label{sec:results}

Having detailed the data and our model in section~\ref{sec:method}, we proceed to obtain point estimates for the parameters of the mixture model. In subsection~\ref{subsec:generator} we study the model performance at the generator level for the two generator choices available, and we include a brief study of detector levels in subsection~\ref{subsec:detector}. All results are reported on a test set which was separated from the train set prior to the model training.

\subsection{Model performance at generator level}\label{subsec:generator}

As detailed in section~\ref{sec:method}, we model the $n_{\mathrm{SD}}$ distribution as originating from a mixture of two Poissonians, which we aim to identify with gluon and quark jets (or, should we want to get rid of perturbative definitions, to operationally define gluon or quark enriched samples). For each choice of hyperparameters, we obtain the Maximum Likelihood Estimates (MLE) of the rates of the Poissonians. $\lambda^{\text{MLE}}_{g}$ and $\lambda^{\text{MLE}}_{q}$, and the mixing fraction between the two $\pi^{\text{MLE}}_{g}$. We define the gluon theme as the theme with the larger rate, as oriented by the perturbative calculations. We obtain the MLE  of the parameters with the help of the Pyro package~\cite{bingham2018pyro,phan2019composable}, which we have verified to coincide with the results obtained through Expectation-Maximization but provide us with a more flexible framework that can incorporate additional features to the generative model and optimize the code appropriately.

\begin{figure}[ht!]
    \centering
    \includegraphics[width=\textwidth]{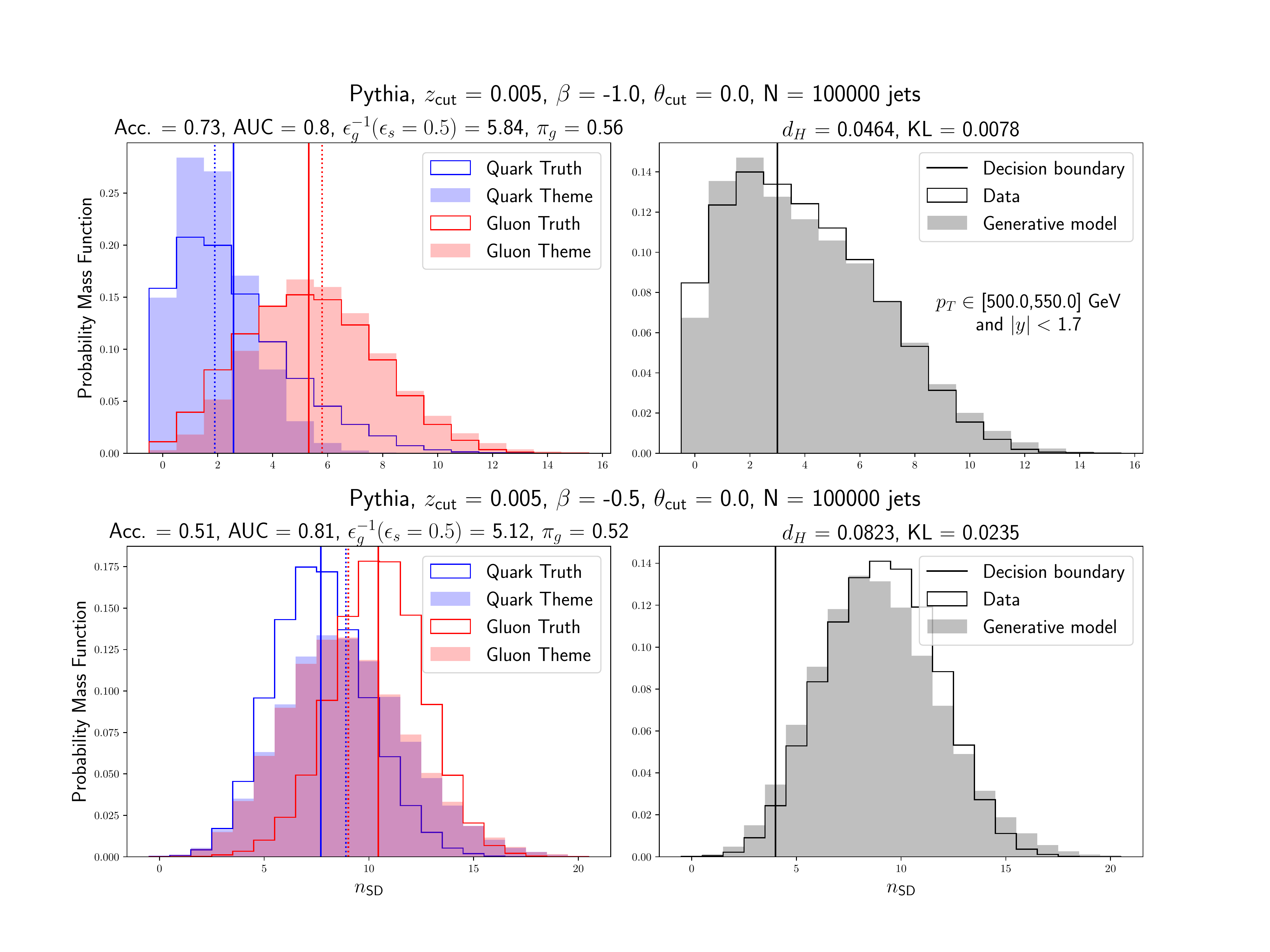}
    \caption{SoftDrop multiplicity distributions for the learned quark and gluon jet themes and the correct ‘true’ answer based on the Pythia generated sample. The upper (lower) row corresponds to a good (bad) choice of hyperparameters.  This can be seen from the supervised side by the accuracy metric and from the unsupervised side by the Hellinger and KL divergence metrics which measure the consistency between the real data and pseudo-data sampled with the learned model parameters.  On the left plots, we show with vertical lines the different Poisson rates while on the right plots we show with a vertical line the threshold corresponding to $p(z=\text{quark}\,|\,n_{\mathrm{SD}})=0.5$. See text for details.}
    \label{fig:mle_params_1}
\end{figure}

For the sake of validation and understanding, we consider three supervised metrics: the accuracy using the probabilistic decision boundary $p(g\,|\,n_{\mathrm{SD}})=p(q\,|\,n_{\mathrm{SD}})$, the inverse gluon mistag rate at 50$\%$ quark efficiency $\epsilon^{-1}_{g}(\epsilon_{q}=50\%)$ and the AUC. Since we are dealing with an unsupervised algorithm, and as discussed above, we also apply the unsupervised metrics defined in section~\ref{sec:method}: the Hellinger distance~\cite{Deza2009} and the Kullback-Leibler divergence. These metrics compare the measured $n_{\mathrm{SD}}$ distribution (without any labels) to the learned total distribution 
\begin{equation}
    p(n_{\mathrm{SD}}\,|\,\text{data})=\pi^{\text{MLE}}_{g}\, \text{Poisson}(\lambda^{\text{MLE}}_{g}) + (1.0-\pi^{\text{MLE}}_{g})\, \text{Poisson}(\lambda^{\text{MLE}}_{q}) .  
\end{equation}
We show two examples of the results we obtain for different SoftDrop hyperparameters in Fig.~\ref{fig:mle_params_1}. In the left column, we show the true underlying distributions and the two learned Poissonians, their respective means and Poissonian rates, and the supervised metrics.  In the right column, we show the data distribution, the learned data distribution and the default decision boundary, along with the unsupervised data-driven metrics.  In the top row, we show a good hyperparameter choice that leads to a data distribution that is well modelled by a mixture of Poissonians, and thus we obtain good supervised and unsupervised metrics. In the bottom row, we show a bad hyperparameter choice leading to a data distribution that is not well modelled by a mixture of Poissonians, and thus we obtain mostly bad supervised and unsupervised metrics, with the exception of the AUC score.

The supervised metrics show that the accuracy and the AUC do not necessarily favour the same models. As shown in Fig.~\ref{fig:mle_params_1}, two very different cases can lead to high AUC, with the accuracy being able to reflect more the true performance of the model. This is due to the fact that the AUC is a more global metric which takes into account every possible threshold in $p(z=\text{quark}\,|\,n_{\mathrm{SD}})$ including the default threshold used for computing the accuracy, $p(z=\text{quark}\,|\,n_{\mathrm{SD}})=0.5$, and can be fooled by moving said threshold. Because the default threshold is theoretically well-motivated, as it takes full advantage of the probabilistic modelling to define a specific boundary between the two classes, it tends better to reflect the goodness of the modelling than the AUC.  As we use probabilistic models for an unsupervised task, interpretability and consistency are important features to keep in mind.  In that sense, the accuracy is more aligned with the unsupervised metrics, which cannot be fooled by moving the decision threshold. The Hellinger distance and the KL divergence see whether the generated dataset is consistent with the measured dataset, taking advantage of the generative procedure.

\begin{figure}[ht!]
    \centering
    \includegraphics[width=\textwidth]{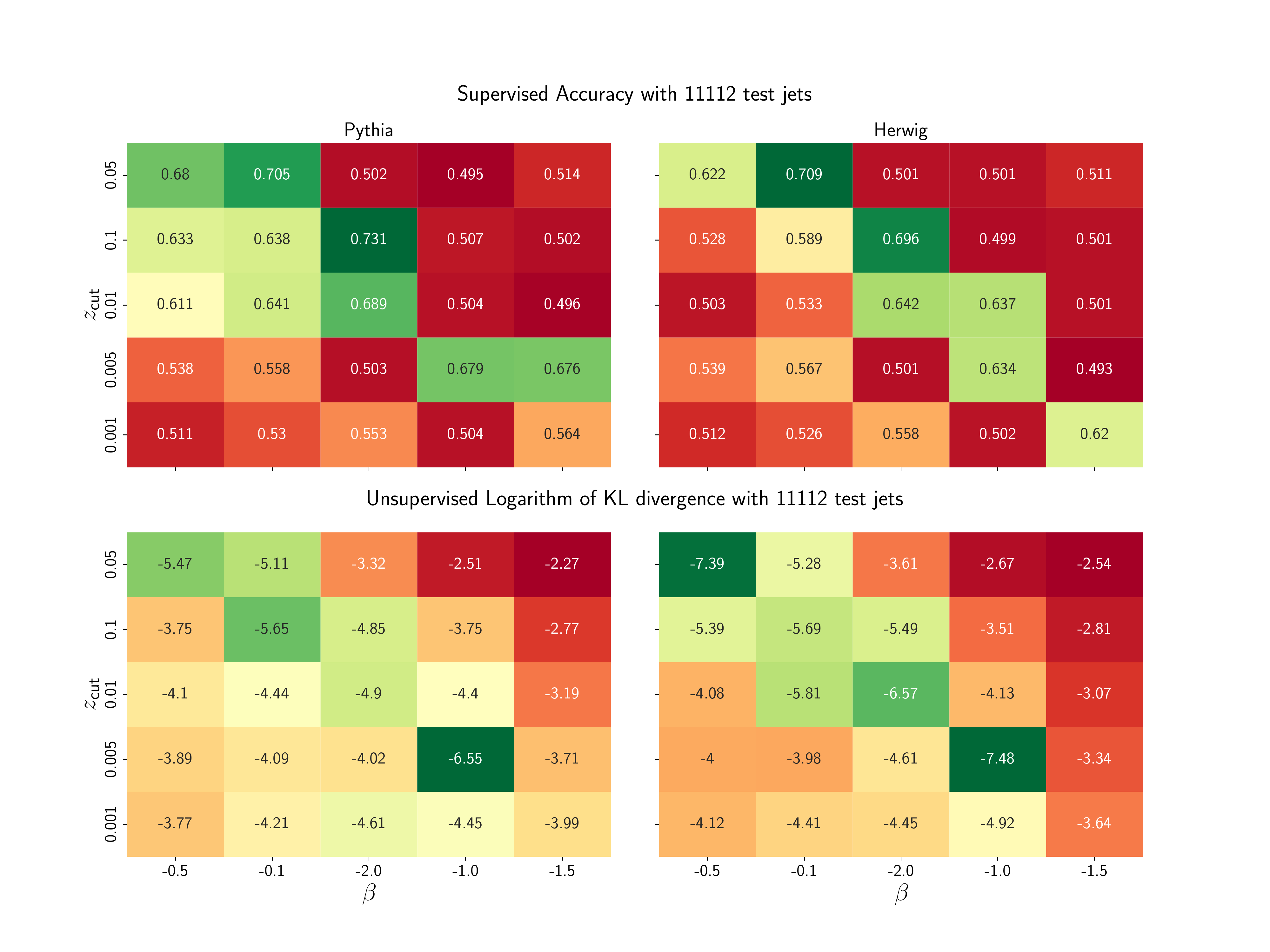}
    \caption{Comparison of supervised and unsupervised learning performance metrics for various hyperparameter choices using the same input data. The color code reflects the goodness of the metrics by coloring in green high accuracies and low KL divergences and vice versa in red. We note that the best hyperparameter choice is consistent with the results reported in Ref.~\cite{Frye:2017yrw}.  Moreover, since there is a fair agreement of the best regions in the upper (supervised) and lower (unsupervised) panels, this suggests that an unsupervised optimization in real data would select a region of good accuracy. Observe that right and left plots correspond not only to different generators, but also to the different setup of the generator parameters.}
    \label{fig:accuracy_and_kl}
\end{figure}

As a next step, we scan the hyperparameter values to study the algorithm performance and how unsupervised metrics can assist us in having a good (unseen) supervised metric. We show the accuracy and the KL divergence for an array of hyperparameter values in  Fig.~\ref{fig:accuracy_and_kl}.  We observe that the accuracy and the KL divergence have a fair agreement in qualifying a good model for a given SoftDrop parameter choice.  Although their respective maximum and minimum do not match exactly, the regions of high accuracy coincide with the regions of low KL divergence. Therefore, we can trust that the accuracy will be increased for a reasonable parameter choice by inspecting the KL divergence and verifying that the obtained quark and gluon themes are suitable. As for the other metrics (not shown in the plot), we find that the Hellinger distance is consistent with the KL divergence and the mistag rate is consistent with the accuracy. The AUC presents the caveat discussed above and thus is less relevant for this study.

From the above results, we see that, by choosing the accuracy as the relevant metric, there is a significant overlap of the good regions in hyperparameter space according to the unsupervised and supervised metrics.  This indicates that classification performance coincides with generative performance, and therefore opens the door for exploring a fully unsupervised approach where the quark/gluon tagging is defined by a relatively simple parameter scan — yielding an unsupervised, interpretable and simple model for classification. 

\begin{figure}[ht!]
    \centering
    \includegraphics[width=\textwidth]{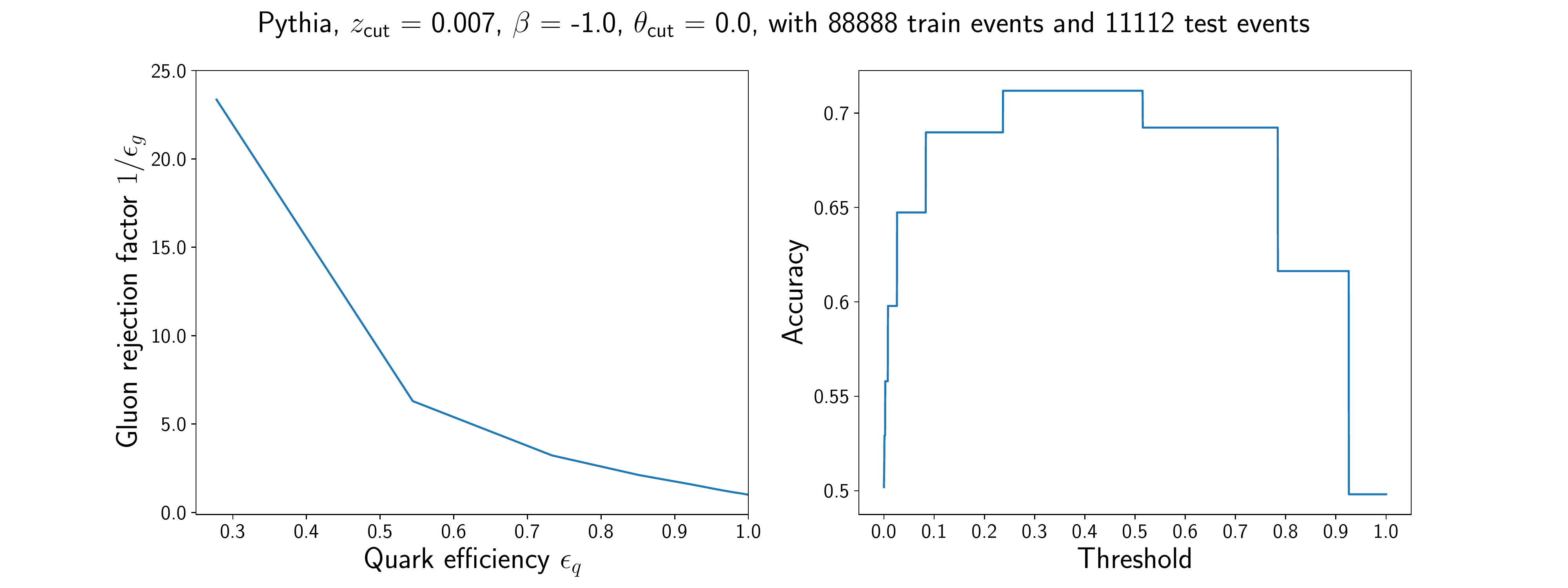}
    \caption{Left: ROC curve for a good hyperparameter choice (AUC=0.77). Right: Accuracy as a function of threshold for the same hyperparameter choice. We observe that the accuracy is constant in regions and that it is maximum in the region that contains the default decision boundary $p(z=\text{quark})=0.5$. The coarse behavior of the accuracy can be traced back to the probabilistic classifier dependence on the discrete $n_{\mathrm{SD}}$. A jet can have a discrete set of $n_{\mathrm{SD}}$ and thus a discrete set of $p(z=\text{quark})$ values.}
    \label{fig:roc_and_accs}
\end{figure}

To study the performance of the proposed unsupervised classifier in some more detail, we compute the ROC and the accuracy as a function of the threshold $c$ to which the classes are defined.  We show in  Fig.~\ref{fig:roc_and_accs} both results for a good point in hyperparameter space.  In a real case scenario, one would only have access to the bottom panel in Fig.~\ref{fig:accuracy_and_kl}, and choosing a point with small KL divergence would yield a tagger that for the threshold $p(z=\text{quark})=0.5$ has an accuracy of roughly $\sim 0.65-0.73$ (Pythia) and $0.62-0.70$ (Herwig).  

\subsection{Detector effects}\label{subsec:detector}
\begin{figure}[ht!]
    \centering
    \includegraphics[width=0.5\textwidth]{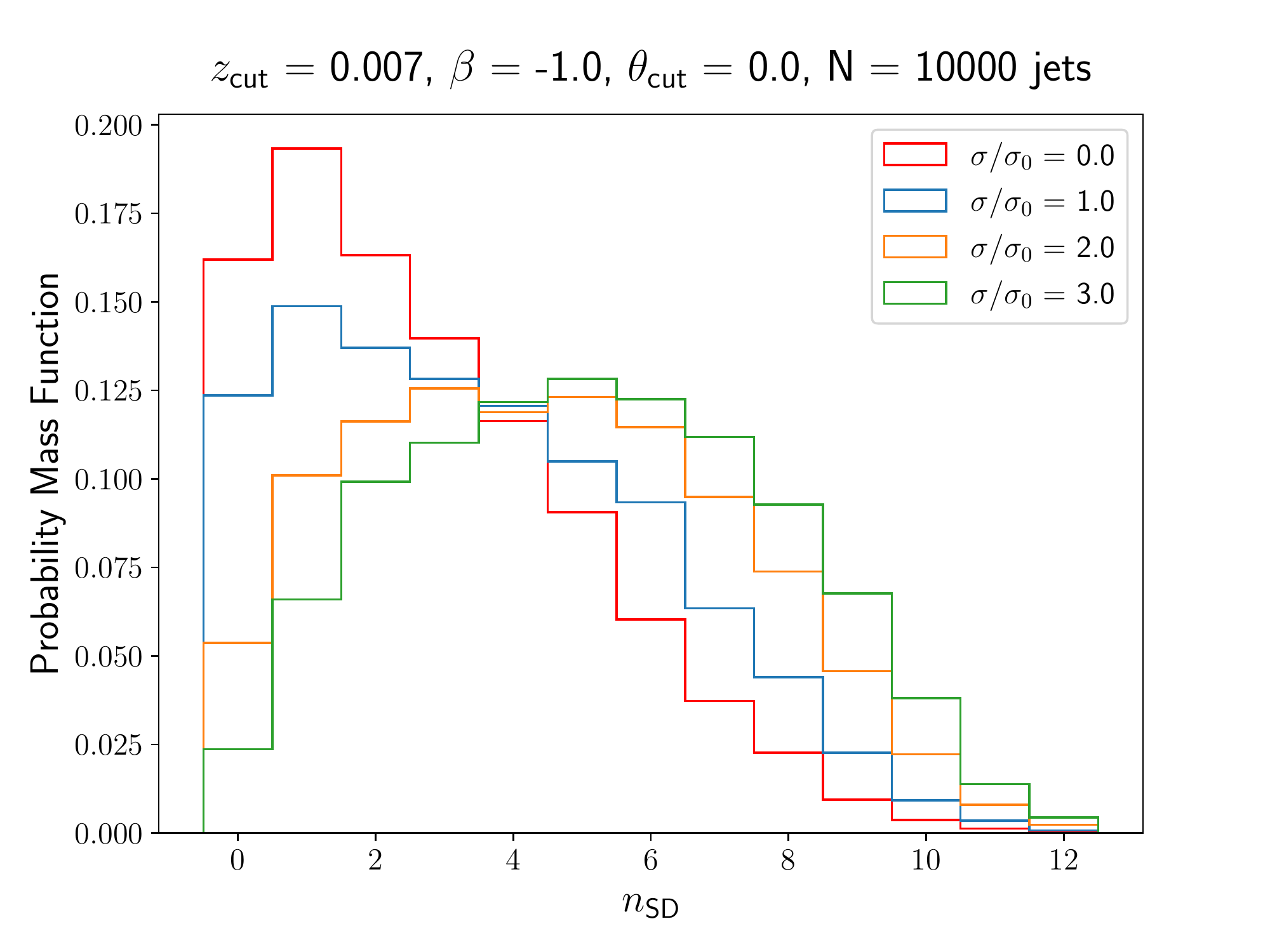}
    \caption{Smeared $n_{\mathrm{SD}}$ distributions obtained by applying the $p_{T}$-dependent emulation of detector effects/response detailed in Eq.~\ref{smearing} and in  Ref.~\cite{Buckley:2019stt} with different scaling factors. A scaling factor of 0 indicates no smearing while a scaling factor of 1 indicates the same smearing factor as in Ref.~\cite{Buckley:2019stt}.}
    \label{fig:det_1}
\end{figure}

\begin{figure}[ht!]
    \centering
    \includegraphics[width=\textwidth]{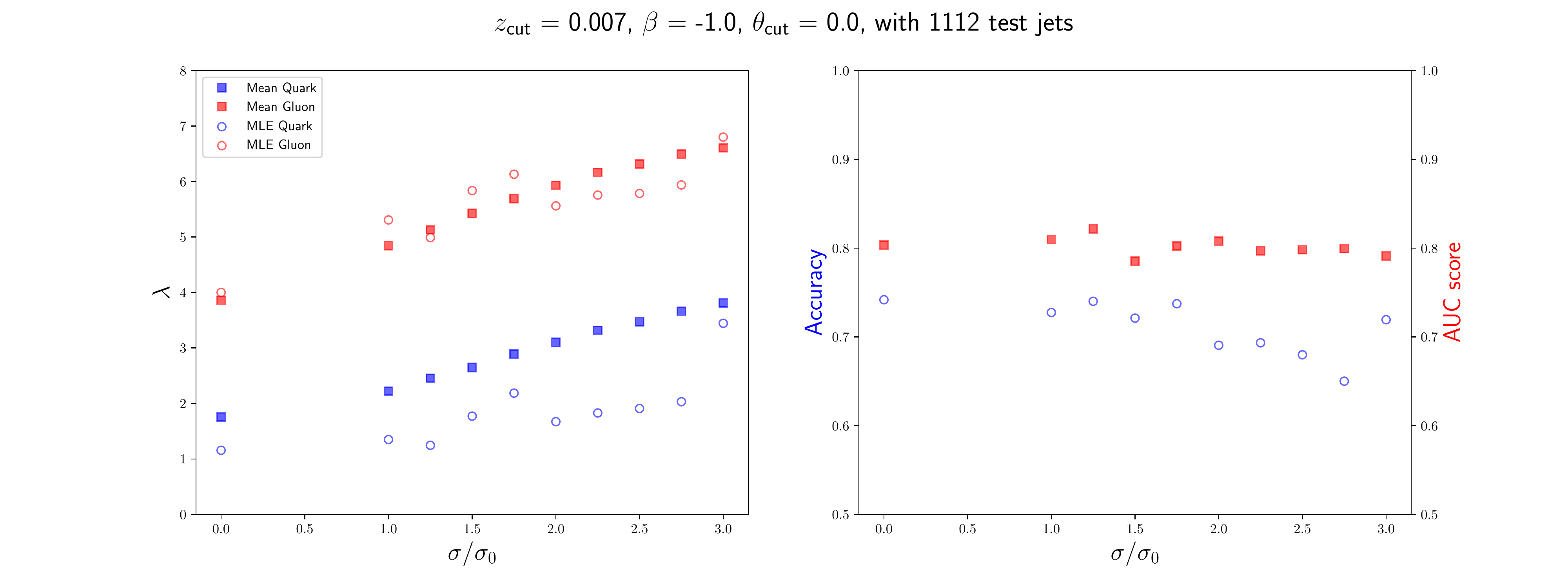}
    \caption{Model performance as a function of the angular smearing. In the left plot we show the obtained Maximum Likelihood Estimates for each Poisson rate and compare them with the means of the true underlying distributions. In the right plot we show the accuracy and the AUC as a function of the scaling factor. }
    \label{fig:det_2}
\end{figure}

To study the sensitivity of our tagger to detector effects, we used the procedure outlined in section 6 in Ref.~\cite{Buckley:2019stt} and smeared the $\eta,\phi$ distribution of each jet constituent. We considered the same smearing as in Ref.~\cite{Buckley:2019stt}, where they spread the $\eta$ and $\phi$ values of a constituent by sampling Gaussian noise with mean zero and standard deviation given by:
\begin{equation}
    \sigma_{0}(p_{T})=\frac{0.028}{1+e^{(p_{T}-25\text{ GeV})/0.1\text{ GeV}}} ,
    \label{smearing}
\end{equation}
where $p_{T}$ is the $p_{T}$ of the constituent. We consider different smearing noise factors $\sigma$ obtained by re-scaling  $\sigma_{0}$ by a global multiplicative factor. The results are shown in Figs.~\ref{fig:det_1} and~\ref{fig:det_2}. Although there is a difference in the MLE due to the change in the distributions, we find no significant alteration in the supervised metrics, and hence in the model performance. It seems that generator effects as simulated are not challenging the model.  This may not be surprising since the model only relies on the assumptions that the integer value $n_{\mathrm{SD}}$ are composed of a mixture of approximately Poissonian distributions. In any case, a more realistic detector simulation should be implemented to verify this analysis, which includes modelling the energy response of various jet constituents, should be implemented. A different and interesting extension is to extend the $|y|$ range to include forward jets. For forward jets, the detector granularity changes and thus the $n_{\mathrm{SD}}$ becomes more dependent on $|y|$. It would then be necessary to introduce similar strategies as the ones detailed for dealing with a large $p_{T}$ range. Finally, we should mention that potential pile-up issues would only have a minor effect on the tagger, not only because $n_\mathrm{SD}$ is a robust observable as it discards soft emission, but also because we are keeping a small radius ($R=0.4$) in the jet clustering algorithm.

\section{Bayesian analysis}\label{sec:bayes}
 
\begin{figure}[ht!]
    \centering
    \includegraphics[width=\textwidth]{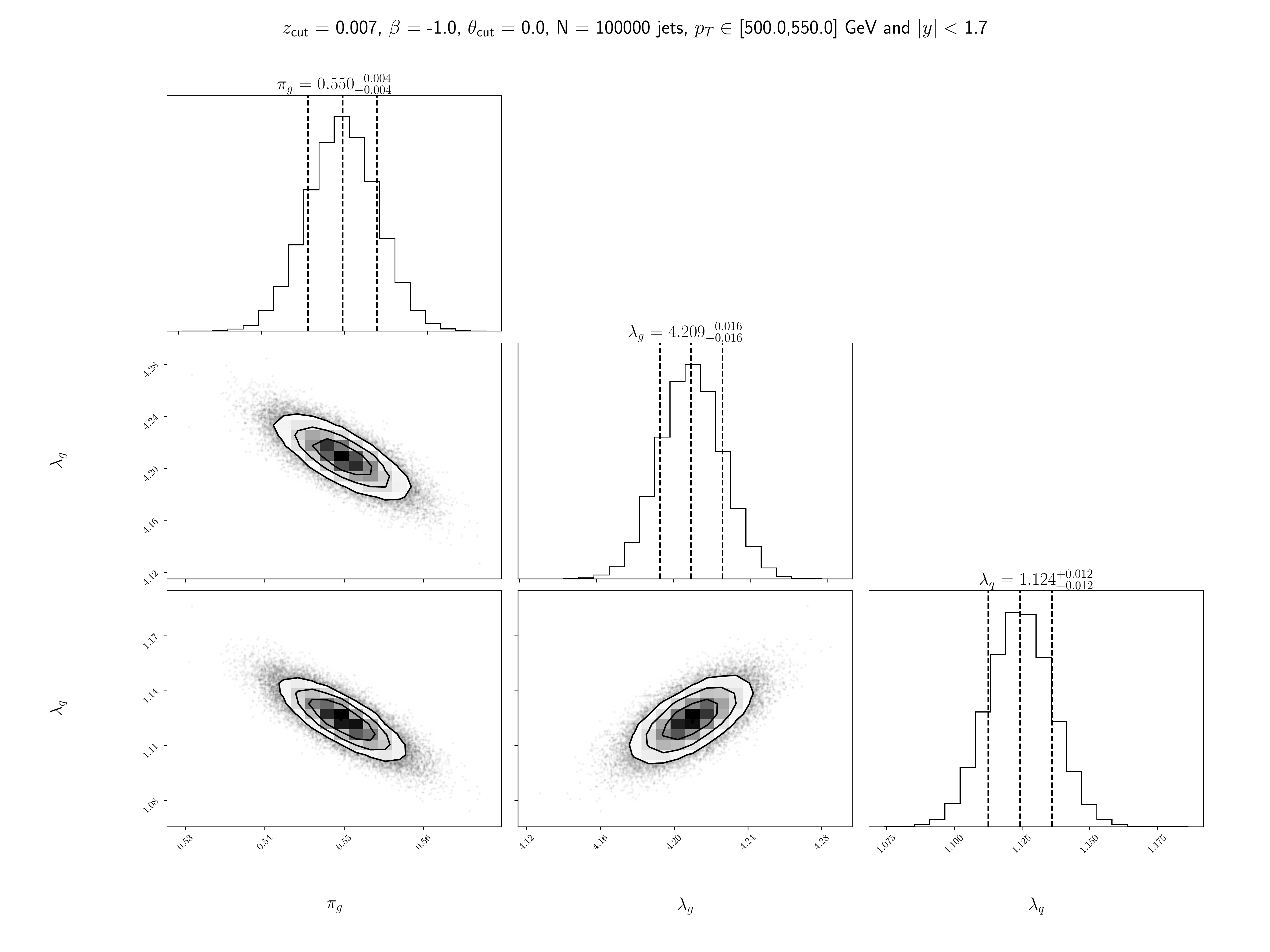}
    \caption{Corner plots for the model parameters $\pi_{g}$, $\lambda_{q}$ and $\lambda_{g}$. The diagonal plots are the 1D marginalized posterior distributions for each parameter while the off-diagonal plots are the pairwise 2D distributions marginalized over the third parameter. The $\pi_{q}$ distribution can be obtained by considering $1-\pi_{g}$. }
    \label{fig:corner_plot}
\end{figure}

\begin{figure}[ht!]
    \centering
    \includegraphics[width=0.5\textwidth]{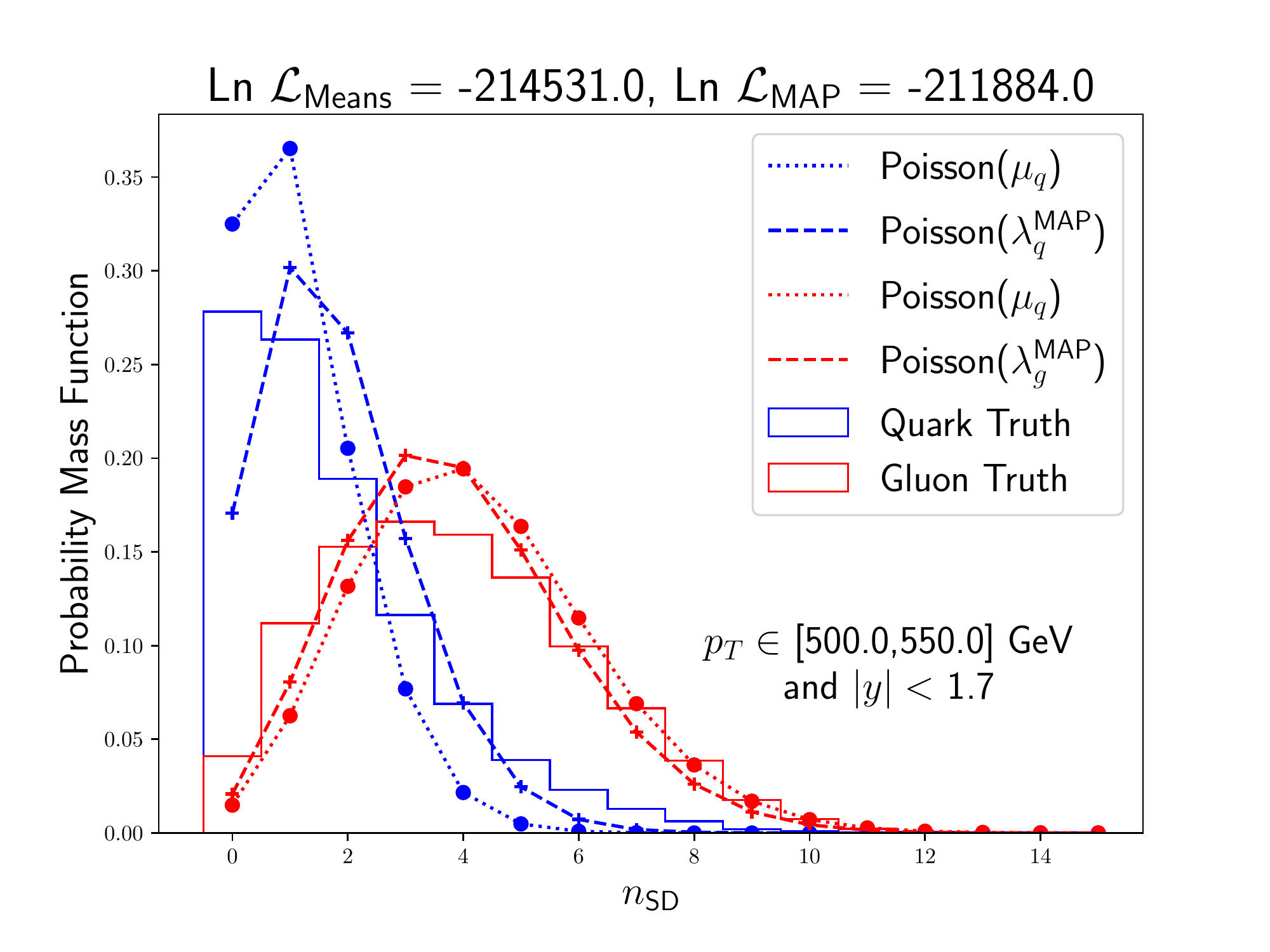}
    \caption{$n_\mathrm{SD}$ distribution comparison between pseudo-data generated through MC (solid) and Poissonians estimates using as rates the mean of the data (dashed) and the Maximum a Posteriori MAP from the Bayesian inference (dashed).  We see that the Poissonian approximations are good enough to distinguish quark from gluon, but there are slight differences when comparing each approximation to its corresponding data.}
    \label{fig:bayes_means_vs_MAPS}
\end{figure}

As a final study for the model, we perform Bayesian inference to obtain the full posterior probability density function over the parameters.  Introducing uniform priors and performing numerical Bayesian inference, we obtain the posterior probabilities of $\pi$ and $\lambda$. In order to achieve this goal, we employ the dedicated {\tt emcee} package~\cite{emcee2013}.  We show the resulting corner plot for a justified hyperparameter choice in Fig.~\ref{fig:corner_plot}. Because we have so many jets and we consider uniform priors, the inference is likelihood dominated with a prominent posterior peak in the MLE.  However, one should not lose sight of the fact that the posterior distribution includes more information than the MAP point estimates since we can quantify the uncertainty of the $\pi$ and $\lambda$ estimation and their correlation. Suppose the generative model for the data is precise enough. In that case, this is a potentially useful application as one could establish a distance between the data-driven posterior distribution and the different Monte Carlo tunes one needs to consider to relate data with Standard Model predictions.   In the specific case of the LL approximation for the $n_\mathrm{SD}$ distribution as Poissonians, we find by inspecting the MC labelled data that the agreement is not good enough to perform such a task (See Fig.~\ref{fig:bayes_means_vs_MAPS}).  However, the approximation is good enough to distinguish the quark from the gluon $n_\mathrm{SD}$ distributions, and therefore to create a good unsupervised classifier. 

\begin{figure}[ht!]
    \centering
    \includegraphics[width=0.5\textwidth]{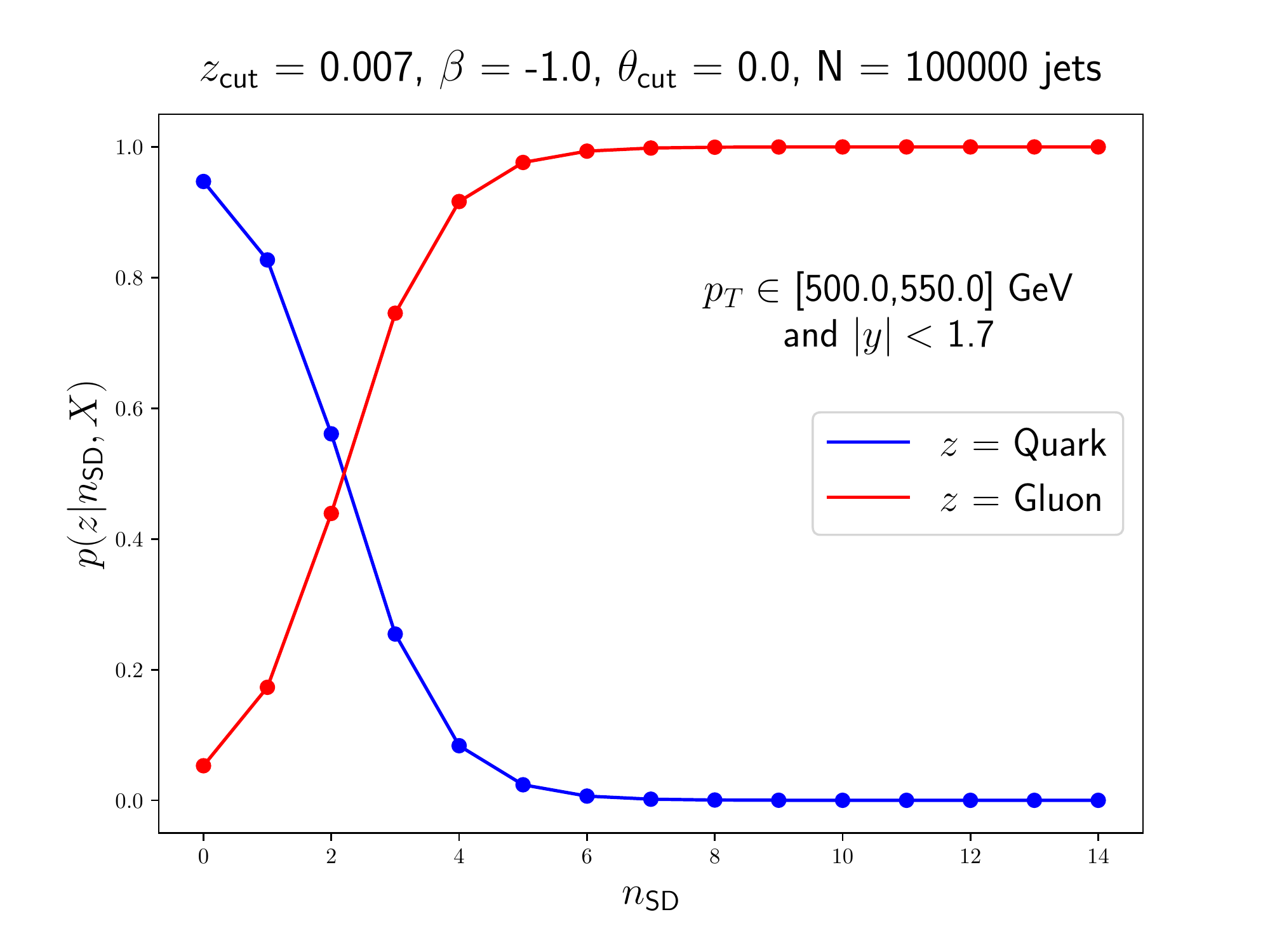}
    \caption{Class assignment probabilities for each $n_{\mathrm{SD}}$ possible value obtained after marginalizing over $\pi$ and $\lambda$. Note that each $n_{\mathrm{SD}}$ has its own probability mass function with two possible outcomes with no constraint arising from summing over $n_{\mathrm{SD}}$.}
    \label{fig:bayes_responsabilities}
\end{figure}

Another feature of Bayesian computation is that we can compute the probability of a given measurement $n_{\mathrm{SD}}$ belonging to class $z$ integrated over the $\lambda_{g}$, $\lambda_{q}$ and $\pi_{g}$ posterior distribution. Using our Monte Carlo samples, we calculate
\begin{equation}
    p(z\,|\,n_{\mathrm{SD}},X) \approx \frac{1}{T}\sum_{t=1}^{T}p(z\,|\,n_{\mathrm{SD}},\pi^{(t)}_{g},\lambda^{(t)}_{g},\lambda^{(t)}_{q})~\,
\end{equation}
where $X$ represents the training dataset and $t$ is the posterior sample index. We show this probability for both classes in Fig.~\ref{fig:bayes_responsabilities}. Although the MLE dominates the likelihood because of our uniform priors and the amount of data, this probability is a more solid estimate when we only care about classifying samples as it considers all possible values of the underlying model parameters weighted by previous measurements through the posterior.  The performance of this tagger using the decision threshold of $p(z=\text{quark})=0.5$ yields an accuracy of $0.71$.

\section{Discussion and outlook}
\label{sec:outlook}
We have proposed an unsupervised data-driven learning algorithm to classify jets induced by quarks or gluons.  The key of the method is to approximate that each class (quark and gluon) has a Poissonian distribution with a different rate for the jets' Soft Drop observable, $n_\mathrm{SD}$.  Therefore, the $n_\mathrm{SD}$ distribution of a sample of an unknown mixture of quark and gluon jets correspond to a mix of Poissonians.  This observation, which is only for approximately constant jet $p_{T}$, allows to set up an unsupervised learning paradigm that can extract the Maximum Likelihood Estimate (MLE) and the posterior distributions for the rate of each Poissonian ($\lambda_q$ and $\lambda_g$) and the fraction of each constituent in the sample ($\pi_{q,g}$) and thus, with this knowledge, one can create a tagger to discriminate quark and gluon induced jets.  This is all achieved without relying on any Monte Carlo generator, nor any previous knowledge other than the assumption that $n_\mathrm{SD}$ is Poisson distributed for each class.  We use the basic principle knowledge that $\lambda_g$ > $\lambda_q$ to assign the tagging of the reconstructed themes.

In the first part of the work, we have defined the generative process of the data according to the above hypothesis and obtained the MLE for the parameters using Stochastic Variational Inference and Expectation-Maximization techniques independently.  We have then designed a quark-gluon tagger and discussed a method to find the best hyperparameters choice for the SoftDrop algorithm that optimise the tagger accuracy.  Since one cannot measure the accuracy in actual data because one does not have access to the labels, we have shown that minimizing the KL divergence between the real data and generated data sampled with the generative model improves the tagger accuracy.  We have verified that the procedure works for different Monte Carlo with different tunes.  One can expect that the described unsupervised tagger can have an accuracy in the range $\approx 0.65 - 0.70$.

We have performed a simple detector effect simulation by smearing the angular coordinate of each jet constituent, and we find that the tagger accuracy remains approximately the same.  This is not surprising since, despite the detector effects, the $n_\mathrm{SD}$ observable is still a counting observable that may vary its value but still be Poissonian distributed with shifted rates.  Therefore the whole machinery of the unsupervised algorithm works essentially the same.

In a second part of the article, we have performed a Bayesian inference on the parameters to extract the full posterior distribution and the correlation between the model parameters, namely $\lambda_q$, $\lambda_g$ and $\pi_g$.  In particular, we have found that the Maximum a Posterior (MAP) approximately coincides with the MLE of the parameters. Furthermore, we have found that although the reconstructed Poissons for each class does not match the labelled data within the posterior uncertainty, the classifier still works quite good.  The reason for this is that, although we can see a slight departure of the approximation of the $n_\mathrm{SD}$ being Poissonian distributed, the two inferred Poissonians for quark and gluon still show a more pronounced difference between them than its corresponding labelled data.  

With the posterior obtained through Bayesian inference, we have designed a quark-gluon tagger based on computing the probability of a jet being induced by either quark or gluon using all the observed data.  This is a more robust tagger since it sees the posterior and hence the correlation between the parameters rather than the point MLE.  With this tagger, we obtain an accuracy of $0.71$.

There are potential improvements and limitations on the proposed algorithm. For example, suppose one could have a model for the $n_\mathrm{SD}$ that goes beyond the LL Poissonian approximation. In that case, one could modify the likelihood and obtain the posterior for the new likelihood parameters.  Although we do not expect this to improve the tagger accuracy considerably, it could help tune a Monte Carlo using unsupervised learning.  If one could have a reliable posterior for specific signal distribution, then one could check whether a Monte Carlo is compatible or not with it.  Observe that, since Monte Carlo generators do not have a handle to set the value for each observable, having a prediction for some observable and its uncertainty provides the necessary information to check whether the Monte Carlo sampling is within the allowed regions defined by the posterior. On other aspects, we have performed simple modelling for the detector effects, which apparently would not affect the tagger performance.  Further investigation in this direction would be helpful to find the actual limitations of the algorithm. 

Finally, we should comment on the challenges that may arise when applying this algorithm in real data. A balanced quark/gluon dataset is far from guaranteed. However, we have verified that the classification and generative powers of the model are robust against a change in the classes fractions up to a 80\% in any class. There is also the possibility of sample contamination with, for example, charm- and bottom-quarks. If there is no need to disentangle charm- and bottom- from light-quarks, then no modification is needed as $n_{\mathrm{SD}}$ is mostly agnostic to quark flavor for relatively fixed jet kinematics. In particular, for jets with $p_T\gg5$ GeV c- and b-jets are as massless as light-jets and they have a similar $n_\mathrm{SD}$ behavior.  If b-tagging and c-tagging is needed, then the model should be extended by incorporating other observables which are sensitive to quark flavor, like the number of displaced vertices in the jet, before searching for three themes instead of two.

Current supervised algorithms to discriminate jets induced by quark or gluon have a non-negligible dependence on Monte Carlo and their tunes, which may hide some intractable systematic uncertainties or biases. Therefore, we find that proposing an unsupervised paradigm for quark-gluon determination is an appealing road that should be transited.  In addition to being interpretable and straightforward, the presented algorithm yields an accuracy in the $0.65-0.7$ range, which is a good achievement for the small number of assumptions on which it relies.

\section*{Acknowledgements} MSz would like to thank the
Jozef Stefan Institute for its warm hospitality during part of this work. 
We thank Referees for relevant and useful suggestions which were included in this revised version.

 \bibliographystyle{JHEP}
 \bibliography{sample}

\providecommand{\href}[2]{#2}\begingroup\raggedright\begin{thebibliography}{10}

\bibitem{Kasieczka:2021xcg}
G.~Kasieczka et~al., {\it {The LHC Olympics 2020: A Community Challenge for
  Anomaly Detection in High Energy Physics}},
  \href{http://arxiv.org/abs/2101.08320}{{\tt arXiv:2101.08320}}.

\bibitem{Aarrestad:2021oeb}
T.~Aarrestad et~al., {\it {The Dark Machines Anomaly Score Challenge: Benchmark
  Data and Model Independent Event Classification for the Large Hadron
  Collider}},  \href{http://arxiv.org/abs/2105.14027}{{\tt arXiv:2105.14027}}.

\bibitem{Choi:2020bnf}
S.~Choi, J.~Lim, and H.~Oh, {\it {Data-driven Estimation of Background
  Distribution through Neural Autoregressive Flows}},
  \href{http://arxiv.org/abs/2008.03636}{{\tt arXiv:2008.03636}}.

\bibitem{Caron:2021wmq}
S.~Caron, L.~Hendriks, and R.~Verheyen, {\it {Rare and Different: Anomaly
  Scores from a combination of likelihood and out-of-distribution models to
  detect new physics at the LHC}},  \href{http://arxiv.org/abs/2106.10164}{{\tt
  arXiv:2106.10164}}.

\bibitem{Dohi:2020eda}
K.~Dohi, {\it {Variational Autoencoders for Jet Simulation}},
  \href{http://arxiv.org/abs/2009.04842}{{\tt arXiv:2009.04842}}.

\bibitem{dAgnolo:2021aun}
R.~T. d'Agnolo, G.~Grosso, M.~Pierini, A.~Wulzer, and M.~Zanetti, {\it
  {Learning New Physics from an Imperfect Machine}},
  \href{http://arxiv.org/abs/2111.13633}{{\tt arXiv:2111.13633}}.

\bibitem{Nachman:2020lpy}
B.~Nachman and D.~Shih, {\it {Anomaly Detection with Density Estimation}},
  \href{http://arxiv.org/abs/2001.04990}{{\tt arXiv:2001.04990}}.

\bibitem{Andreassen:2020nkr}
A.~Andreassen, B.~Nachman, and D.~Shih, {\it {Simulation Assisted
  Likelihood-free Anomaly Detection}},
  \href{http://arxiv.org/abs/2001.05001}{{\tt arXiv:2001.05001}}.

\bibitem{Hajer2020}
J.~Hajer, Y.-Y. Li, T.~Liu, and H.~Wang, {\it Novelty detection meets collider
  physics},  {\em Physical Review D} {\bf 101} (Apr, 2020).

\bibitem{roy2020robust}
T.~S. Roy and A.~H. Vijay, {\it A robust anomaly finder based on autoencoders},
   2020.

\bibitem{Andreassen:2018apy}
A.~Andreassen, I.~Feige, C.~Frye, and M.~D. Schwartz, {\it {JUNIPR: a Framework
  for Unsupervised Machine Learning in Particle Physics}},  {\em Eur. Phys. J.
  C} {\bf 79} (2019), no.~2 102, [\href{http://arxiv.org/abs/1804.09720}{{\tt
  arXiv:1804.09720}}].

\bibitem{CMS:2014jvv}
{\bf CMS} Collaboration, V.~Khachatryan et~al., {\it {Search for dark matter,
  extra dimensions, and unparticles in monojet events in
  proton\textendash{}proton collisions at $\sqrt{s} = 8$ TeV}},  {\em Eur.
  Phys. J. C} {\bf 75} (2015), no.~5 235,
  [\href{http://arxiv.org/abs/1408.3583}{{\tt arXiv:1408.3583}}].

\bibitem{Dokshitzer:1991he}
Y.~L. Dokshitzer, V.~A. Khoze, and T.~Sjostrand, {\it {Rapidity gaps in Higgs
  production}},  {\em Phys. Lett. B} {\bf 274} (1992) 116--121.

\bibitem{Rainwater:1998kj}
D.~L. Rainwater, D.~Zeppenfeld, and K.~Hagiwara, {\it {Searching for
  $H\to\tau^+\tau^-$ in weak boson fusion at the CERN LHC}},  {\em Phys. Rev.
  D} {\bf 59} (1998) 014037, [\href{http://arxiv.org/abs/hep-ph/9808468}{{\tt
  hep-ph/9808468}}].

\bibitem{Bhattacherjee:2016bpy}
B.~Bhattacherjee, S.~Mukhopadhyay, M.~M. Nojiri, Y.~Sakaki, and B.~R. Webber,
  {\it {Quark-gluon discrimination in the search for gluino pair production at
  the LHC}},  {\em JHEP} {\bf 01} (2017) 044,
  [\href{http://arxiv.org/abs/1609.08781}{{\tt arXiv:1609.08781}}].

\bibitem{Gallicchio:2011xq}
J.~Gallicchio and M.~D. Schwartz, {\it {Quark and Gluon Tagging at the LHC}},
  {\em Phys. Rev. Lett.} {\bf 107} (2011) 172001,
  [\href{http://arxiv.org/abs/1106.3076}{{\tt arXiv:1106.3076}}].

\bibitem{Larkoski:2013eya}
A.~J. Larkoski, G.~P. Salam, and J.~Thaler, {\it {Energy Correlation Functions
  for Jet Substructure}},  {\em JHEP} {\bf 06} (2013) 108,
  [\href{http://arxiv.org/abs/1305.0007}{{\tt arXiv:1305.0007}}].

\bibitem{Larkoski:2014pca}
A.~J. Larkoski, J.~Thaler, and W.~J. Waalewijn, {\it {Gaining (Mutual)
  Information about Quark/Gluon Discrimination}},  {\em JHEP} {\bf 11} (2014)
  129, [\href{http://arxiv.org/abs/1408.3122}{{\tt arXiv:1408.3122}}].

\bibitem{Bhattacherjee:2015psa}
B.~Bhattacherjee, S.~Mukhopadhyay, M.~M. Nojiri, Y.~Sakaki, and B.~R. Webber,
  {\it {Associated jet and subjet rates in light-quark and gluon jet
  discrimination}},  {\em JHEP} {\bf 04} (2015) 131,
  [\href{http://arxiv.org/abs/1501.04794}{{\tt arXiv:1501.04794}}].

\bibitem{FerreiradeLima:2016gcz}
D.~Ferreira~de Lima, P.~Petrov, D.~Soper, and M.~Spannowsky, {\it {Quark-Gluon
  tagging with Shower Deconstruction: Unearthing dark matter and Higgs
  couplings}},  {\em Phys. Rev. D} {\bf 95} (2017), no.~3 034001,
  [\href{http://arxiv.org/abs/1607.06031}{{\tt arXiv:1607.06031}}].

\bibitem{Kasieczka:2018lwf}
G.~Kasieczka, N.~Kiefer, T.~Plehn, and J.~M. Thompson, {\it {Quark-Gluon
  Tagging: Machine Learning vs Detector}},  {\em SciPost Phys.} {\bf 6} (2019),
  no.~6 069, [\href{http://arxiv.org/abs/1812.09223}{{\tt arXiv:1812.09223}}].

\bibitem{ATLAS:2014vax}
{\bf ATLAS} Collaboration, G.~Aad et~al., {\it {Light-quark and gluon jet
  discrimination in $pp$ collisions at $\sqrt{s}=7\mathrm {\ TeV}$ with the
  ATLAS detector}},  {\em Eur. Phys. J. C} {\bf 74} (2014), no.~8 3023,
  [\href{http://arxiv.org/abs/1405.6583}{{\tt arXiv:1405.6583}}].

\bibitem{CMS:2013kfa}
{\bf CMS} Collaboration, {\it {Performance of quark/gluon discrimination in 8
  TeV pp data}}, .

\bibitem{Larkoski:2014wba}
A.~J. Larkoski, S.~Marzani, G.~Soyez, and J.~Thaler, {\it {Soft Drop}},  {\em
  JHEP} {\bf 05} (2014) 146, [\href{http://arxiv.org/abs/1402.2657}{{\tt
  arXiv:1402.2657}}].

\bibitem{Frye:2017yrw}
C.~Frye, A.~J. Larkoski, J.~Thaler, and K.~Zhou, {\it {Casimir Meets Poisson:
  Improved Quark/Gluon Discrimination with Counting Observables}},  {\em JHEP}
  {\bf 09} (2017) 083, [\href{http://arxiv.org/abs/1704.06266}{{\tt
  arXiv:1704.06266}}].

\bibitem{Komiske:2018cqr}
P.~T. Komiske, E.~M. Metodiev, and J.~Thaler, {\it {Energy Flow Networks: Deep
  Sets for Particle Jets}},  {\em JHEP} {\bf 01} (2019) 121,
  [\href{http://arxiv.org/abs/1810.05165}{{\tt arXiv:1810.05165}}].

\bibitem{Zenodo:EnergyFlow:Pythia8QGs}
P.~Komiske, E.~Metodiev, and J.~Thaler, {\it Pythia8 quark and gluon jets for
  energy flow},  {\em Zenodo} (2019).

\bibitem{Sjostrand:2014zea}
T.~Sj\"ostrand, S.~Ask, J.~R. Christiansen, R.~Corke, N.~Desai, P.~Ilten,
  S.~Mrenna, S.~Prestel, C.~O. Rasmussen, and P.~Z. Skands, {\it {An
  introduction to PYTHIA 8.2}},  {\em Comput. Phys. Commun.} {\bf 191} (2015)
  159--177, [\href{http://arxiv.org/abs/1410.3012}{{\tt arXiv:1410.3012}}].

\bibitem{Zenodo:EnergyFlow:Herwig7QGs}
A.~Pathak, P.~Komiske, E.~Metodiev, and M.~Schwartz, {\it Herwig7.1 quark and
  gluon jets},  {\em Zenodo} (2019).

\bibitem{Bahr:2008pv}
M.~Bahr et~al., {\it {Herwig++ Physics and Manual}},  {\em Eur. Phys. J. C}
  {\bf 58} (2008) 639--707, [\href{http://arxiv.org/abs/0803.0883}{{\tt
  arXiv:0803.0883}}].

\bibitem{Bellm:2015jjp}
J.~Bellm et~al., {\it {Herwig 7.0/Herwig++ 3.0 release note}},  {\em Eur. Phys.
  J. C} {\bf 76} (2016), no.~4 196,
  [\href{http://arxiv.org/abs/1512.01178}{{\tt arXiv:1512.01178}}].

\bibitem{Komiske:2018vkc}
P.~T. Komiske, E.~M. Metodiev, and J.~Thaler, {\it {An operational definition
  of quark and gluon jets}},  {\em JHEP} {\bf 11} (2018) 059,
  [\href{http://arxiv.org/abs/1809.01140}{{\tt arXiv:1809.01140}}].

\bibitem{Dokshitzer:1997in}
Y.~L. Dokshitzer, G.~D. Leder, S.~Moretti, and B.~R. Webber, {\it {Better jet
  clustering algorithms}},  {\em JHEP} {\bf 08} (1997) 001,
  [\href{http://arxiv.org/abs/hep-ph/9707323}{{\tt hep-ph/9707323}}].

\bibitem{Metodiev:2017vrx}
E.~M. Metodiev, B.~Nachman, and J.~Thaler, {\it {Classification without labels:
  Learning from mixed samples in high energy physics}},  {\em JHEP} {\bf 10}
  (2017) 174, [\href{http://arxiv.org/abs/1708.02949}{{\tt arXiv:1708.02949}}].

\bibitem{Komiske:2018oaa}
P.~T. Komiske, E.~M. Metodiev, B.~Nachman, and M.~D. Schwartz, {\it {Learning
  to classify from impure samples with high-dimensional data}},  {\em Phys.
  Rev.} {\bf D98} (2018), no.~1 011502,
  [\href{http://arxiv.org/abs/1801.10158}{{\tt arXiv:1801.10158}}].

\bibitem{Metodiev:2018ftz}
E.~M. Metodiev and J.~Thaler, {\it {Jet Topics: Disentangling Quarks and Gluons
  at Colliders}},  {\em Phys. Rev. Lett.} {\bf 120} (2018), no.~24 241602,
  [\href{http://arxiv.org/abs/1802.00008}{{\tt arXiv:1802.00008}}].

\bibitem{Dillon:2019cqt}
B.~M. Dillon, D.~A. Faroughy, and J.~F. Kamenik, {\it {Uncovering latent jet
  substructure}},  {\em Phys. Rev.} {\bf D100} (2019), no.~5 056002,
  [\href{http://arxiv.org/abs/1904.04200}{{\tt arXiv:1904.04200}}].

\bibitem{Alvarez:2019knh}
E.~Alvarez, F.~Lamagna, and M.~Szewc, {\it {Topic Model for four-top at the
  LHC}},  {\em JHEP} {\bf 01} (2020) 049,
  [\href{http://arxiv.org/abs/1911.09699}{{\tt arXiv:1911.09699}}].
  [JHEP20,049(2020)].

\bibitem{Dillon:2020quc}
B.~M. Dillon, D.~A. Faroughy, J.~F. Kamenik, and M.~Szewc, {\it {Learning the
  latent structure of collider events}},  {\em JHEP} {\bf 10} (2020) 206,
  [\href{http://arxiv.org/abs/2005.12319}{{\tt arXiv:2005.12319}}].

\bibitem{Alvarez:2021hxu}
E.~Alvarez, B.~M. Dillon, D.~A. Faroughy, J.~F. Kamenik, F.~Lamagna, and
  M.~Szewc, {\it {Bayesian Probabilistic Modelling for Four-Tops at the LHC}},
  \href{http://arxiv.org/abs/2107.00668}{{\tt arXiv:2107.00668}}.

\bibitem{Graziani:2021vai}
G.~Graziani, L.~Anderlini, S.~Mariani, E.~Franzoso, L.~Pappalardo, and
  P.~di~Nezza, {\it {A Neural-Network-defined Gaussian Mixture Model for
  particle identification applied to the LHCb fixed-target programme}},
  \href{http://arxiv.org/abs/2110.10259}{{\tt arXiv:2110.10259}}.

\bibitem{_t_p_nek_2015}
M.~{\v{S}}t{\v{e}}p{\'{a}}nek, J.~Franc, and V.~K{\r{u}}s, {\it Modification of
  gaussian mixture models for data classification in high energy physics},
  {\em Journal of Physics: Conference Series} {\bf 574} (jan, 2015) 012150.

\bibitem{Dillon:2021aeo}
B.~M. Dillon, D.~A. Faroughy, J.~F. Kamenik, and M.~Szewc, {\it {Learning
  Latent Jet Structure}},  {\em Symmetry} {\bf 13} (2021), no.~7 1167.

\bibitem{celeux:hal-01961077}
G.~Celeux, S.~Fr{\"u}hwirth-Schnatter, and C.~Robert, {\it {Model Selection for
  Mixture Models-Perspectives and Strategies}},  in {\em {Handbook of Mixture
  Analysis}}.
\newblock {CRC Press}, Dec., 2018.

\bibitem{bishop}
C.~M. Bishop, {\em Pattern Recognition and Machine Learning}.
\newblock Springer Berlin Heidelberg, Berlin, Heidelberg, 2013.

\bibitem{bingham2018pyro}
E.~Bingham, J.~P. Chen, M.~Jankowiak, F.~Obermeyer, N.~Pradhan, T.~Karaletsos,
  R.~Singh, P.~Szerlip, P.~Horsfall, and N.~D. Goodman, {\it {Pyro: Deep
  Universal Probabilistic Programming}},  {\em Journal of Machine Learning
  Research} (2018).

\bibitem{phan2019composable}
D.~Phan, N.~Pradhan, and M.~Jankowiak, {\it Composable effects for flexible and
  accelerated probabilistic programming in numpyro},
  \href{http://arxiv.org/abs/1912.11554}{{\tt arXiv:1912.11554}}.

\bibitem{Deza2009}
M.~M. Deza and E.~Deza, {\em Encyclopedia of Distances}, pp.~1--583.
\newblock Springer Berlin Heidelberg, Berlin, Heidelberg, 2009.

\bibitem{Komiske:2019fks}
P.~T. Komiske, E.~M. Metodiev, and J.~Thaler, {\it {Metric Space of Collider
  Events}},  {\em Phys. Rev. Lett.} {\bf 123} (2019), no.~4 041801,
  [\href{http://arxiv.org/abs/1902.02346}{{\tt arXiv:1902.02346}}].

\bibitem{Buckley:2019stt}
A.~Buckley, D.~Kar, and K.~Nordstr\"om, {\it {Fast simulation of detector
  effects in Rivet}},  {\em SciPost Phys.} {\bf 8} (2020) 025,
  [\href{http://arxiv.org/abs/1910.01637}{{\tt arXiv:1910.01637}}].

\bibitem{emcee2013}
D.~Foreman-Mackey, D.~W. Hogg, D.~Lang, and J.~Goodman, {\it emcee: The mcmc
  hammer},  {\em Publications of the Astronomical Society of the Pacific} {\bf
  125} (03, 2013) 306–312.

\end{thebibliography}\endgroup

\end{document}